\documentstyle[12pt]{article}
\newcommand{\ze}{\raisebox{-0.4em}{$\scriptstyle {0}$}}

\newcommand{\Te}{\raisebox{-0.4em}{$\scriptstyle {T}$}}

\newcounter{Nsec}
\newcounter{multieqs}

\begin{document}
\baselineskip=26pt plus2pt minus2pt

\begin{center}
{\Large \bf MAXWELL ELECTROMAGNETIC THEORY,\\
PLANCK'S RADIATION LAW AND\\[0.3cm]
BOSE-EINSTEIN STATISTICS} \\[0.3cm]
(November, 1995)
\end{center}

\vspace{0.3cm}

\begin{center}
{\bf H.M. Fran\c{c}a}\footnote{Instituto de F\'{\i}sica, Universidade
de S\~ao Paulo, C.P. 66318, 05389-970 S\~ao Paulo, SP, Brazil.}, {\bf
A. Maia Jr.}\footnote{Instituto de Matem\'atica, Estat\'{\i}stica e
Ci\^encia da Computa\c{c}\~ao, Universidade Estadual de Cam\-pi\-nas,
C.P. 6065, 13081-970 Campinas, SP, Brazil.} and {\bf C.P. Malta}$^1$
\end{center}

\baselineskip=26pt plus2pt minus2pt

\vspace{1.5cm}

\begin{center}
{\bf ABSTRACT}
\end{center}

We give an example in which it is possible to understand quantum
statistics using classical concepts. This is done by studying the
interaction of charged matter oscillators with the thermal and
zeropoint electromagnetic fields characteristic of quantum
electro\-dynamics and classical stochastic electrodynamics. Planck's
formula for the spectral distribution and the elements of energy $\,
\hbar\omega\,$ are interpreted without resorting to discontinuities. We
also show the aspects in which our model calculation complement other
derivations of blackbody radiation spectrum without quantum
assumptions.

%\vspace{1.3cm}

%\noindent PACS numbers: \ 03.50.De, 03.65Bz, 05.40+j

\vfill\eject

Recently, Tersoff and Bayer$^{\textstyle [1]}$ have shown that the
statistics obeyed by the particles cannot be used as a criterion for their
distinguishability. Here we show that the general proposal of Tersoff
and Bayer, when adapted to the blackbody radiation phenomenon, provides
a clue for bringing quantum and classical physics into a closer relation.
In this regard, the similarity between quantum electrodynamics (QED)
and the classical theory called stochastic electrodynamics (SED) has
been stressed by several authors in recent years$^{\textstyle
[2\mbox{--}7]}$. It should be mentioned that to obtain a closer
relation between the classical and quantum approaches was one of the
goals of Planck and Einstein research in the first and second decades
of this century. It is well known that Einstein himself never gave up
to obtain a deeper understanding of the light-quantum concept within
the realm of classical electrodynamics$^{\textstyle [8]}$. Following
then the proposal of Tersoff and Bayer, we show that it is possible to
reconcile the statistics of Bose and Einstein with classical
electrodynamics. As a natural byproduct of our analysis, we interpret
Planck's radiation law for the blackbody radiation, without using
concepts strange to classical physics, like discontinuities in the
energy of the charged particle during the emission or absorption
processes. This reinforces the attempts to interpret the ``photon''
model using only the undulatory aspect of the electro\-magnetic field,
thus providing a simpler picture of the ``light-quantum''$^{\textstyle
[9]}$.

For a {\em single\/} one-dimensional nonrelativistic matter oscillator
(mass \ $m \, $, \ charge \ $e$ \ and frequency \ $\omega$ \ such that
\ $e^2\omega/mc^3 \ll 1$) interacting with a cavity radiation with
spectral distribution \ $\rho(\omega, T) = \rho \ze (\omega) +
\rho\Te(\omega) \, $, \ there is a simple relation (valid within both
QED and SED) between the average oscillator energy \ $\langle \epsilon
\rangle$ \ and \ $\rho(\omega,T) \, $, \ namely$^{\textstyle [10]}$
\vspace{0.3cm}
\begin{eqnarray}
\langle \epsilon \rangle & = &
   \frac{4\pi}{3} \; \frac{e^2}{m} \; \omega^2
    \hspace{-0.7cm}\int^{\hspace{5ex}
	 \omega_{\rm max}}_0\limits \hspace{-5mm}
   \frac{d\omega' \left[ \rho \ze(\omega')+\rho \Te(\omega')
	      \right]}{\left({\omega'}^2-\omega^2 \right)^2 + \left(
    {\displaystyle \frac{2}{3} \; \frac{e^2}{mc^3} \; {\omega'}^3}
	  	\right)^2} \; \; \cong \; \; \frac{\pi^2c^3}{\omega^2} \;
       \rho(\omega,T)  \; \; = \nonumber \\[0.3cm]
& = & \frac{\hbar \omega}{2} + \frac{\pi^2c^3}{\omega^2} \;
      \rho \Te (\omega) \; \; \equiv \; \; \frac{\hbar\omega}{2}
      + \langle u \rangle   \quad ,
\end{eqnarray} %(1)

\vspace{0.3cm}

\noindent where \ $\omega_{\rm max}$ \ is the maximum frequency
compatible with nonrelativistic motion.

In (1) we have considered that \ $\hbar\omega/2$ \ is the average
zeropoint (or zero temperature) energy of the oscillator, $\langle u
\rangle$ \ is its average thermal energy and
\vspace{0.3cm}
\begin{equation}
\rho \ze (\omega) \; \; = \; \; \frac{\hbar \omega^3}{2\pi^2c^3}
     \quad ,
\end{equation}%(2)

\vspace{0.3cm}

\noindent is the spectral distribution of the zeropoint electromagnetic
field (which exists in both SED and QED). We are denoting by \ $\rho\Te
(\omega)$ \ the spectral distribution of thermal radiation with
temperature \ $T \, $.  \ As far as we know, formula (1) was firstly
obtained by Planck (see ref.~[8], p.~870) with \ $\rho \ze (\omega) = 0
\, $. \ The inclusion of the zeropoint field contribution to the
oscillator average energy was made later, in 1911, also by
Planck$^{\textstyle [7]}$.

According to SED, the spectral distribution \ $\rho(\omega,T)$ \ which
appears in (1), is associated to the {\em random fluctuations\/} of the
classical electromagnetic fields existing inside the cavity. These
fields are correlated, and one of the correlation functions is defined
through the ensemble average$^{\textstyle [3,4]}$
\vspace{0.3cm}
\begin{equation}
\frac{\langle \vec E(t) \cdot \vec E(0) \rangle}{4\pi} \; \; = \; \;
     \hspace{-0.7cm}\int^{\kern 5ex \omega_{\rm max}}_0\limits
		 \hspace{-5mm} d\omega \, \rho(\omega,T) \,
     \cos(\omega t) \quad ,
\end{equation} %(3)

\vspace{0.3cm}

\noindent where \ $\vec E(t)$ \ is the electric field at time $\,t\,$,
in the origin of coordinate system (oscillator position). It is also
assumed that \ $\langle \vec E (t) \rangle = 0$. We shall adopt the
classical stochastic electrodynamics approach in what follows.

Within the realm of this enlarged version of Maxwell's electromagnetic
theory it is possible to derive the exact expression for \
$\rho(\omega, T)$ \ without quantum assumptions. This is well known and
a number of researchers have published, and commented, several
(different) derivations many times in the past$^{\textstyle [4]}$. To
our knowledge the first one was published by Einstein and
Stern$^{\textstyle [11]}$ in the paper entitled ``Some Arguments for
the Assumptions of a Molecular Agitation at Absolute Zero'' published
in 1913. According to Einstein and Stern, ``the assumption of the
zeropoint energy opens a way for deriving Planck's radiation formula
without resorting to any discontinuities whatsoever''.

For completeness we shall outline here another simple classical
derivation which is based on the interaction of a {\em single\/}
oscillator with cavity radiation. According to this derivation, which
was discussed previously by some authors$^{\textstyle [4,12]}$, the
starting point is the Einstein formula for the thermal fluctuations of
the oscillator energy. In agreement with the Einstein formula, which
may be interpreted as definition of temperature$^{\textstyle [13]}$
(see also ref.~[8] p.~874), the total oscillator energy has thermal and
zeropoint fluctuations. Therefore, it is possible to show that the
energy variance is such that$^{\textstyle [13]}$
\vspace{0.3cm}
\begin{equation}
\langle \epsilon^2 \rangle - \langle \epsilon \rangle^2 \; \; = \; \;
     \langle \epsilon \rangle^2 \; \; = \; \;
     kT^2 \; \frac{\partial}{\partial T} \; \langle \epsilon \rangle +
     \left( \frac{\hbar\omega}{2} \right)^2  \quad ,
\end{equation} %(4)

\vspace{0.3cm}

\noindent where \ $\langle \epsilon \rangle$ \ is given by (1). As we
already said, this result was used previously by several authors (see
for instance Boyer$^{\textstyle [13]}$ or Fran\c{c}a and
Santos$^{\textstyle [12]}$). In equation (4), the first term of the
last equality represents the thermal fluctuations. The second term is
associated to the zeropoint fluctuations of the oscillator energy. It
is assumed that the thermal and zeropoint fluctuations are
statistically independent (see de la Pe\~na$^{\textstyle [4]}$, p.~475)
in the sense that their variances simply add$^{\textstyle [14]}$, as is
shown in equation~(4).

The differential equation (4) has an exact solution for the average
energy \ $\langle \epsilon \rangle$ \ of the oscillator as a function
of the temperature $\, T\,$. \ This solution is precisely$^{\textstyle
[13]}$
\vspace{0.3cm}
\begin{equation}
\langle \epsilon \rangle  \; \; = \; \; \frac{\hbar\omega}{2} \;
   \coth\left( \frac{\hbar\omega}{2kT} \right) \; \; = \; \;
    \frac{\hbar\omega}{2} + \frac{\hbar\omega}{\exp\left(
      {\displaystyle \frac{\hbar\omega}{kT}} \right) - 1}
       \quad ,
\end{equation} %(5)

\vspace{0.3cm}

\noindent as it is easy to verify. The limit \ $kT \gg \hbar\omega$ \
gives the expected result, namely \linebreak $\langle \epsilon \rangle
\, \simeq \, \langle u \rangle \, \simeq \, kT\,$, \ and the low
temperature limit is also correct. Besides these results, one can
obtain the spectral distribution \ $\rho (\omega, T)$, by combining (5)
and (1).  To our knowledge this is one of the simplest classical
derivations of the Planck spectral distribution. An important remark is
that the result (5) follows directly from (1) and (4), and the
equations (1) and (4) follows from the classical stochastic equations
of motion for the charged oscillator$^{\textstyle [2\mbox{--}4]}$.

In order to make a closer comparison with the original 1900 Planck's
paper, and to make the connection with the statistics of Planck,
Bose$^{\textstyle [15]}$ and Einstein, it is necessary to study the
same physical system considered by Planck or a similar one. We may
consider, for instance, a macroscopic group of $A$ identical matter
oscillators, which are coupled to each other like the ions in a solid
lattice. We assume, for instance, that these oscillators are in the
walls of a cavity which exist in a solid material. The cavity has
thermal and zeropoint radiation, which are in equilibrium with the
matter oscillators of the surrounding walls.  Therefore, this physical
model will be naturally related to the Einstein and Debye models for
the thermal vibrations of a solid body.

A classical interpretation of the Debye law for the specific heat of
solids was presented by Blanco, Fran\c{c}a and Santos$^{\textstyle
[16]}$ quite recently. These authors have used the SED approach, thus
presenting a derivation of the Debye specific-heat law, using classical
equations for Brownian ions coupled by linear forces. Their work
``provides a classical interpretation of the specific heat as being due
to a {\em continuous\/} distribution of energies of the normal-modes
vibrations''. The spectral distribution of these vibrations is given by
\ $\rho(\omega, T) = (\hbar\omega^3/2\pi^2 c^3)
\coth(\hbar\omega/2kT)$. Therefore, this work$^{\textstyle [16]}$ shows
``that the close connection (discovered by Einstein in 1907) between
the Planck blackbody spectrum and the specific-heat law is maintained
even if both are interpreted along classical ideas''. We recommend the
paper by Blanco et al.$^{\textstyle [16]}$ to the interested reader.

Here, however, we shall consider a simpler system (the Planck
original system), consisting of $A$ identical {\em uncoupled\/} matter
oscillators. That is, we shall assume that the oscillators are
separated from each other.  Consequently, each oscillator interacts
only with the electromagnetic waves associated to the thermal and
the zeropoint radiations. In this sense, our model is physically equivalent
to a rarefied gas of polarizable molecules in equilibrium with
blackbody radiation. Analyzing this simple system (uncoupled
oscillators plus radiation) along the ideas proposed by Tersoff and
Bayer, we show that it is possible to understand better the relation
between Maxwell electromagnetic theory, Planck's radiation law and
Bose-Einstein statistics. In this sense our paper complements the other
derivations (classical and quantum) of the blackbody radiation spectrum.

Let us denote the {\em thermal\/} energy of the $A$ oscillators by $U$.
Therefore, it is possible to write $U$ as
\vspace{0.3cm}
\begin{equation}
U \; \; = \; \; \sum^A_{i=1} u_i \quad ,
\end{equation} %(6)

\vspace{0.3cm}

\noindent where \ $u_i$ \ is only the {\em thermal\/} part of the total
energy of the $i$-th oscillator. The index $\,i\,$ denotes the site,
or position, of the oscillator in space. The zeropoint energy of the
oscillators is not included in (6).

On physical grounds, in order for the temperature $\,T\,$ of a system
of $A$ oscillators to be meaningful, we expect the system to be in
equilibrium, the average total thermal energy \ $\langle U \rangle$ \
of the oscillators to be constant and we also expect the system to be
very large ($A \gg 1$ \ or \ $A \rightarrow \infty$). According to (1)
and (6) the average thermal energy is simply:
\begin{equation}
\langle U \rangle \; \; = \; \; A \langle u \rangle \quad ,
\end{equation}  %(7)
with \ $\langle U \rangle \simeq 0$, if \ $T \rightarrow 0$, or \
$\langle U \rangle \simeq A\,kT$ \ if \ $kT \gg \hbar\omega\,$. \
However, the actual energy $\,U\,$ is not constant. It has {\em
fluctuations\/}, due to energy exchange with the zeropoint and the thermal
radiations, as well as due to the flow of energy in and out of the box,
or volume, containing the oscillators and the electromagnetic
radiation.  Therefore, we expect $\, U\,$ to be a {\em random\/}
variable which varies {\em continuously\/} within some interval, namely
\begin{equation}
U_{\rm min} \;\; \leq \; \; U \; \; \leq \; \; U_{\rm max} \quad ,
\end{equation} %(8)
where \ $U_{\rm min}$ \ is close to \ $U_{\rm max}$ \ for a large system
($A \rightarrow \infty$). Nevertheless, we shall assume that \ $U_{\rm
max} > U_{\rm min}$ \ and that \ $U_{\rm min} \rightarrow 0$ \ for very
low temperatures ($T \rightarrow 0$). Moreover, we expect the
fluctuations in the thermal energy $\, U\,$ to be more important for
small $T$. However, the total thermal and zeropoint energy of the
system (oscillators plus radiation) is assumed to be conserved, via
energy conservation laws discussed in the recent article by
Cole$^{\textstyle [17]}$, for instance.

We need to introduce one important step in order to apply the Tersoff
and Bayer's method and to connect it with the quantum picture and the early
Planck's paper. So, similarly to Planck, let us {\em formally\/} divide the
total thermal energy $\, U\,$ of the matter oscillators into  fractions
$\,q\,$,
\begin{equation}
q \; \; \equiv \; \; \frac{U}{N} \quad ,
\end{equation} %(9)

\noindent  $N$ \ being an arbitrary integer that is (at least) of the same
order
of magnitude as \ $ A $. This subdivision of the thermal energy \ $U$ \ is
necessary because we want to {\em count\/} the number of different ways one can
distribute a {\em continuous\/} amount of thermal energy \ $U$ \
amongst $\,A\,$ oscillators.

Since the oscillator thermal energies can take {\em arbitrary\/} values
\ $u_i$ \ such that (6) is valid, we can introduce a set of numbers \
$\{ n_i\} \,$, \ {\em not necessarily integers\/} (this is an striking
departure from the Planck's 1900 paper), such that
\begin{equation}
u_i  \; \equiv  \; n_i \, q  \quad ,
\end{equation} %(10)
and
\begin{equation}
N \; \equiv \; \sum^A_{i=1} \, n_i \quad .
\end{equation} %(11)

\vspace{0.3cm}

Of course the possibility of integer numbers \ $\{ n_i \}$ \ is also
compatible with (10) and (11).  In any case, each \ $n_i$ \ represents
the amount of fractions \ $q$ \ which an oscillator has.  We shall
assume that these fractions of energy, $q=U/N$, are {\em
distinguishable\/} within the context of Tersoff and Bayer classical
approach. Moreover, each fraction $\,q\,$ varies continuously and has
fluctuations (see (8) and (9)) and, therefore, {\em cannot be strictly
identified with the hypothetical Planck's quantum}.

The approach of Tersoff and Bayer requires the assignment of different
(random) probability weights \ $\alpha_i$ \ (for the absorption of
thermal energy) to identical oscillators ($i = 1,2,\ldots A$).  So, in
order to use their approach, an essential step is to justify this
assumption. This can be achieved if we recognize that the probability
weights \ $\alpha_i$ \ must be related to the probability of having
some thermal energy available to each oscillator in its position in
space. Regarding this point, our approach will be similar to that
proposed by Ehrenfest$^{\textstyle [18]}$ in 1911.  Ehrenfest
generalized Boltzmann's concept of ``a priori probability'' and used
probability weights which vary with the energy interval. In the
following we shall present a simple model which makes an explicit use
of this idea.

Let us assume that the probability weight \ $\alpha_i$ \ is
proportional to the energy density available in site $i$, which is the
$i$-th oscillator position. As was explained before (see (3)), the
total electric field, associated to the thermal and zeropoint
radiation, in that point is a {\em random\/} vector such that
\begin{equation}
 \langle \vec E_i (t) \rangle \; = \; 0
     \qquad \qquad \mbox{and}  \qquad \qquad
\vec E_i^2 (t) \; \geq \; 0 \quad .
\end{equation} %(12)

Therefore, it is quite natural to assume that the probability weights\
$\alpha_i$ \ may vary according to the following simple model:
\vspace{0.3cm}
\begin{equation}
d\alpha_i \; = \; \mbox{const} \times p\left( \vec E_i \right)
   \, d\!\left( \frac{\vec E_i^2}{4\pi} \right)     \quad ,
\end{equation} %(13)

\vspace{0.3cm}

\noindent where \ $p(\vec E_i)$ \ is the probability distribution of
the classical, Gaussian, random field, characteristic of SED (see
Boyer$^{\textstyle [3]}$ and Milonni$^{\textstyle [7]}$, chapters 2 and
8). In other words, according to our model, the probability weights \
$\alpha_i$ \ may increase (or decrease) with the electromagnetic energy
density available in the oscillator position. We shall, therefore,
assume that each \ $\alpha_i$ \ may vary randomly from zero (no energy
available) to one, namely
\begin{equation}
0 \; \leq \; \alpha_i \; \leq \; 1  \quad ,
\end{equation} %(14)
as is expected on physical grounds.

At this point it is also important to recall that, according to Tersoff
and Bayer, ``it is arguably more natural (i.e., entails a weaker
assumption) to take the probability weights ($\alpha_i$) as arbitrary
and random than equal''. We shall also assume that
\vspace{0.3cm}
\begin{equation}
\sum^A_{i=1} \alpha_i \; = \; 1
\end{equation} %(15)

\vspace{0.3cm}

\noindent within our model. This is required since the entire system
(oscillators plus radiation) is in equilibrium. Intuitively speaking,
{\em the flow of electromagnetic energy\/} from one oscillator site to
the other oscillators sites, is compatible with our model equations
(13), (14) and (15). Moreover, the explicit knowledge of the functional
form of the probability weights \ $\{ \alpha_i \}$ \ is not necessary
in order to apply the Tersoff and Bayer method as we shall see in a
moment.

The above considerations will allow us to calculate the probability \
$P\{n_i\} \, $, \ associated to the numbers \ $\{n_i\}$ \ of {\em
distinguishable\/} fractions $\,(q = U/N)$, which are distributed
amongst \ $A$ \ {\em distinguishable\/} oscillators according to (9),
(10) and (11). This probability, as proposed by Tersoff and Bayer, is
\vspace{0.3cm}
\begin{equation}
P\{n_i\} \; \; = \; \; N! \, \left\langle \, \prod^A_{i=1} \;
     \frac{\alpha_i^{n_i}}{n_i!} \, \right\rangle \; \; = \; \;
    N! \; \prod^A_{i=1} \left(\int^1_0 d\alpha_i \;
     \frac{\alpha_i^{n_i}}{n_i!} \, \right) \, \delta \left(
      1 - \sum^A_{j=1} \alpha_j \right) \quad ,
\end{equation} %(16)

\vspace{0.3cm}

\noindent where the ensemble average \ $\langle \ \rangle$ \ is related
to the continuous and random variation of the probability weights \ $\{
\alpha_i \} \,$. \ In (16) \ $n_i!\,$ is the gamma function \
$\Gamma(n_i+1)$ \ if \ $n_i$ \ is not an integer.

The exact result for this expression, valid even in the general case
in which the numbers \ $n_i$ \ are {\em not integers}, is
simply$^{\textstyle [1]}$
\vspace{0.3cm}
\begin{equation}
P\{n_i\} \; \; = \; \; \frac{N! \, (A-1)!}{(N+A-1)!}  \quad ,
\end{equation} %(17)

\vspace{0.3cm}

\noindent independently of the numbers \ $\{n_i\}\,$. \ This remarkable
result of Tersoff and Bayer is clearly valid within the realm of SED.
It allows the derivation of Planck's radiation law without resorting to
any discontinuities in the emission or absorption of radiation by the
matter oscillators, and to interpret Planck's counting (or statistical)
method$^{\textstyle [8]}$ using only the classical properties of
electromagnetic fields and mechanical oscillators.

As  \ $A$ \ and \ $N$ \ are assumed to be very large
numbers ($A\gg 1$ \ and \ $N\gg 1$), using the Stirling
approximation and the Boltzmann relation
between the entropy and probability, the entropy {\em per oscillator\/}
is given by
\begin{equation}
S^* \; \; \equiv \; \; - \; \frac{k}{A} \; \ln \, P\{n_i\} \; \;
    \cong \; \; k \left[ \left( 1+\frac{N}{A} \right) \, \ln \left(
     1+ \frac{N}{A} \right) - \frac{N}{A}  \, \ln \, \frac{N}{A} \,
      \right]   \ \  .
\end{equation} %(18)
Using (9), the entropy \ $S^*$ can be expressed in terms of the variable \ $U/q
A$, namely,
\begin{equation}
S^* \; \;
    \cong \; \; k \left[ \left( 1+\frac{U}{q A} \right) \, \ln \left(
     1+ \frac{U}{q A} \right) - \frac{U}{q A}  \, \ln \, \frac{U}{q A} \,
      \right]   \ \  .
\end{equation} %(19)

We shall call this function \ $S^*$ the {\em probabilistic\/}
entropy per oscillator. It will be identified with the {\em caloric\/}
entropy \ $S$ \ in the {\em thermodynamic\/} limit \ $A \rightarrow
\infty$. It should be remarked that \ $S$ has to be a function of $\omega/T$,
according to the classical Wien law$^{\textstyle [7]}$.

Since $\, S^*\,$ has to be finite, it is expected that the ratio \ $U/q A$
will be finite in the thermodynamic limit. We shall show that the ratio \ $U/q
A$ \  will be related
to the properties of a single oscillator, with thermal energy \
$\langle \epsilon \rangle - {\displaystyle \frac{\hbar\omega}{2}}\,$,
in accordance with the SED calculation (see equation (5)).

The relation between the temperature \ $T$ \ and the caloric entropy is
well known and can be written as \ $1/T = \partial S/\partial\langle
\epsilon \rangle\,$. Therefore, according to SED we have
\vspace{0.3cm}
\begin{equation}
\frac{1}{T} \; \; = \; \; \frac{\partial S}{\partial\langle \epsilon
  \rangle} \; \; = \; \; \frac{k}{\hbar\omega} \; \ln\left( 1 +
   \frac{\hbar\omega}{\langle \epsilon \rangle - {\displaystyle
	 \frac{\hbar\omega}{2}}} \right)   \quad ,
\end{equation} %(20)

\vspace{0.3cm}

\noindent where the last equality in (20) was obtained from our
previous equation (5). Therefore, the simple integration of (20) will
lead to
\vspace{0.3cm}
\begin{equation}
S \; = \; k \left[ \left( 1 +
  \frac{\langle u \rangle}{\hbar\omega} \right)
   \ln \left( 1 + \frac{\langle u \rangle}{\hbar\omega} \right) -
     \left( \frac{\langle u \rangle}{\hbar\omega} \right) \ln\left(
       \frac{\langle u \rangle}{\hbar\omega}\right) \right] \quad ,
\end{equation} %(21)

\vspace{0.3cm}

\noindent  where \ $\langle u \rangle = \langle \epsilon \rangle -
\hbar \omega/2$.

Consequently, in the thermodynamic limit  $A \rightarrow \infty$,
one must have
\vspace{0.3cm}
\begin{equation}
S^* \; \; \longrightarrow \; \; S  \quad .
\end{equation} %(22)

\vspace{0.3cm}

According to (6), (7) and (8) the ratio \ $q A/U$ \ can be written as
\vspace{0.3cm}
\begin{equation}
 q \; \frac{A}{U} \; \; = \; \; q \left(
  \frac{A}{\langle U \rangle + \delta U}  \right) \; \; = \; \;
   \frac{q}{\langle u \rangle +{\displaystyle \frac{\delta U}{A}}}
     \quad  ,
\end{equation}  %(23)

\vspace{0.3cm}

\noindent because \ $U \equiv A\langle u \rangle + \delta U$ \ with \
$\langle \delta U \rangle = 0$ \ and \ $|\delta U| \sim \sqrt{A} \,
\langle u \rangle\,$ for large $\,A$.

Therefore, comparing the expressions (19) and (21) and taking into
account (23), the limiting process indicated in (22) is equivalent to
\vspace{0.3cm}
\begin{equation}
\frac{q}{\langle u \rangle + {\displaystyle \frac{\delta U}{A}}} \; \;
   \longrightarrow  \; \; \frac{\hbar\omega}{\langle u \rangle}
      \quad ,
\end{equation}%(24)
or
\begin{equation}
q \; \; \longrightarrow \; \; \hbar\omega \left( 1 +
   \frac{\delta U}{A\langle u \rangle} \right) \; = \;
      \hbar\omega \; \frac{U}{\langle U \rangle}   \quad .
\end{equation} %(25)

\vspace{0.3cm}

\noindent In other words, $\,\hbar\omega$ \ is the {\em average\/}
value, $\, \langle q \rangle$, \ of the energy fractions used here in
order to apply the Tersoff and Bayer counting method. As we have
already mentioned, these fractions $\, q\,$ cannot be strictly
identified with the hypothetical Planck's quantum $\, \hbar \omega$.
The result (25) for $\, q\,$ is a consequence of the counting method of
Tersoff and Bayer applied to the thermal energy $\, U\,$ which varies
{\em continuously\/} and has {\em random fluctuations}. Moreover, one
can show that the value $\, \hbar\omega\,$ obtained for $\, \langle q
\rangle\,$ can be traced back to the assumed existence of zeropoint
electromagnetic fluctuations with average energy $\,
\hbar\omega/2 \,$ per mode (see (1), (2), (5), (20) and (21)).

Similar conclusion was obtained by Barranco and Fran\c{c}a$^{\textstyle
[9]}$ in a recent paper, although through a different reasoning. These
authors replaced the Einstein concept of ``random spontaneous
emission'' (see also Milonni~[4] p.~63) by the concept of ``stimulated
emission by the random zeropoint fields''. Using this new concept and
the old result by Einstein and Ehrenfest$^{\textstyle [9]}$, namely,
the energy of the molecule (which interacts with the radiation fields)
can vary {\em continuously}, Barranco and Fran\c{c}a extended the
Einstein (1917) and Einstein-Ehrenfest (1923) works to the realm of
classical stochastic electrodynamics. As a result, Compton's kinematic
relations were interpreted within the realm of a classical theory, that
is, treating the molecules as classical particles and the radiation as
classical waves.

We have seen that, according to SED, the thermal ``photons'' can be
described using only the classical aspects of the radiation field.
This simplified view of thermal radiation is important because it is
also well known that the photoelectric and the Compton effects can be
explained without the particle model of the ``photon''$^{\textstyle
[19]}$.  These facts are known since the pioneering works by Richardson
(1914), Wentzel (1927), Schr\"odinger (1927) and Klein and Nishina
(1929) among many others$^{\textstyle [20]}$. This suggests to us that
the particle model of the ``photon'' is not firmly established and that
QED essentially provides the rules of the interaction without
explicitly invoking the corpuscular character of the
``photon''$^{\textstyle [19,20]}$ (see also Milonni$^{\textstyle [7]}$,
pg.~19). It is also important to comment another ``evidence'' of the
``corpuscular aspect'' of the ``photon'' (i.e., its
``indivisibility''). This was offered by the ``almost'' 100\%
anticorrelation affecting the detection of the ``single-photon'' (i.e.
an experiment with a very weak light beam) by two detectors placed on
the two output modes of a beam-splitter$^{\textstyle [21]}$. However,
the results of this experiment are not conclusive because they may be
compatible with classical wave models of light$^{\textstyle [22]}$.

Nevertheless, we want to remark that the Tersoff and Bayer derivation
of quantum statistics based on non-traditional assumptions does not disprove
the validity of the conventional ones, and that those writers wisely
caution that: ``Since we have shown that either set of counting
assumptions can lead to quantum statistics, the statistics which
``particles'' obey cannot be a criterion for their
distinguishability''. In this regard, there are some indications that
the distinguishability of ``photons'' (or classical electromagnetic
pulses) interacting with optical instruments has been experimentally
verified. This important discovery, which was reported by at least two
different groups very recently$^{\textstyle [23]}$, has not received
the attention it deserves.

Another remark is that blackbody derivations are notorious as being
feasible on a number of different models. Therefore, we would advise
that a multiplicity of derivations from different points of view merely
indicates that a successful derivation (conventional or
non-traditional) cannot constitute a proof of any assumed model. In
this regard we would like to mention other attempts to obtain a
classical derivation of Planck's radiation law. As far as we know,
Einstein and Stern$^{\textstyle [11]}$ in 1913, and Nernst$^{\textstyle
[24]}$ in 1916 were the first physicists who claimed to have derived
Planck's formula without assuming discontinuities (see also ref.~[7]).
More recently Boyer$^{\textstyle [25\mbox{--}27]}$ and
others$^{\textstyle [28\mbox{--}33]}$ have proposed different ways to
derive the blackbody radiation spectrum without quantum assumptions.
These approaches were based on Planck's second theory (1911) which
introduced the concept of zeropoint energy of the electro\-magnetic
field. It is worthwhile to recall that nowadays these zeropoint
fluctuations have been extensively observed experimentally$^{\textstyle
[4,7,34\mbox{--}36]}$.

According to A. Hermann$^{\textstyle [37]}$, ``Planck realized the
radical consequences of his formula only much later, and for many
years he made attempts to harmonize the quantum of action with
classical theory. Planck did not believe that he made a complete break
with the past. He gave the impression that it was impossible to speak
of a counted probability without a subdivision of energy''. This is in
agreement with Ehrenfest old opinion that the elements of energy in
Planck's calculation were only a {\em formal device\/}: ``The
permutation of the energy symbols ($\hbar\omega$) had no more physical
significance than the permutation of the divider symbols'' (see
ref.~[18] pp.~255 and 256). Ehrenfest's formal proof of Planck's
combinatorial formula (our equation~(17)) was enclosed in a letter to
Lorentz in 1914. In another (historical) letter addressed to R.W. Wood
in 1931 (see ref.~[37] p.~23 for the full reproduction), Planck
describes in detail the psychological aspects of his approach.
``Briefly summarized, what I did can be described as simply an act of
desperation. I was ready to sacrifice every one of my previous
convictions about physical laws.  Boltzmann had explained how
thermodynamic equilibrium is established by means of a statistical
equilibrium, and if such an approach is applied to the equilibrium
between matter and radiation, one finds that the continuous loss of
energy into radiation can be prevented by assuming that energy is
forced, at the outset, to remain together in certain quanta. This was a
purely formal assumption and I really did not give it much thought
except that, no matter what the cost, I must bring about a positive
result''.

According to SED and QED the inevitable loss of energy into radiation
is naturally prevented by the vacuum zeropoint electromagnetic field,
which acts as an energy reservoir$^{\textstyle [2\mbox{--}7]}$. On the
other hand, the Tersoff and Bayer approach, adapted here to classical
stochastic electrodynamics and the old blackbody radiation problem,
also clarifies (on classical grounds) the statistical method used by
Planck. Therefore, Planck's formulas (see our equations (5), (17),(19),
(20), (21) and (25)) do not {\em necessarily\/} represent a complete
break with classical physics and the continuity principle. Moreover, in
their interesting paper, Tersoff and Bayer suggest that ``quantum
mechanics may be after all compatible with classical ideas of locality
and distinguishability; the key appears to lie in a reevaluation of the
underlying assumptions about probabilities''. This is precisely the
approach used in SED. In our opinion this general idea deserves further
attention.

\vspace{1cm}
%\pagebreak

\noindent{\bf ACKNOWLEDGMENT}
\vspace{0.3cm}

We want to thank the financial support from Funda\c{c}\~ao de Amparo
\`a Pesquisa do Estado de S\~ao Paulo (FAPESP) and Conselho Nacional de
Desenvolvimento Cient\'{\i}fico e Tecnol\'ogico (CNPq). We also
acknowledge M. Cattani, M. Ferrero, E.C. Lima Filho, T.W.  Marshall and
E. Santos for a critical reading of the manuscript and valuable
comments.

%\vspace{1.5cm}

\baselineskip=24.5pt plus2pt minus2pt

\vfill\eject

\noindent{\bf REFERENCES}
\begin{list}{}{\setlength{\leftmargin}{7mm}\labelwidth2.5cm
\itemsep0pt \parsep0pt}

\item[{[1]}] J. Tersoff and D. Bayer, ``Quantum statistics for
distinguishable particles'', {\sl Phys. Rev. Lett.} {\bf 50}, 553
(1983).

\item[{[2]}] H.M. Fran\c{c}a and T.W. Marshall, ``Excited states in
stochastic electrodynamics'', {\sl Phys. Rev.} {\bf A38}, 3258 (1988).
See also K. Dechoum and H.M. Fran\c{c}a, ``Non-Heisenberg states of the
harmonic oscillator'', {\sl Found. Phys.} {\bf 25}, 1599 (1995).

\item[{[3]}] T.W. Marshall, ``Random electrodynamics'', {\sl Proc. R.
Soc.} (London) {\bf 276A}, 475 (1963). See also T.H. Boyer, ``General
connection between random electrodynamics and quantum electrodynamics
for free electromagnetic fields and for dipole oscillator systems'',
{\sl Phys. Rev.} {\bf D11}, 809 (1975) and T.H. Boyer, in {\sl
``Foundations of Radiation Theory and Quantum Electrodynamics''},
edited by A.O. Barut (Plenum, New York, 1980), pp.~49--63.

\item[{[4]}] L. de la Pe\~na, ``Stochastic electrodynamics: its
development, present situation and perspectives'', in {\sl ``Stochastic
Processes Applied to Physics and Other Related Fields''}, eds.: B.
Gomes, S.M. Moore, A.M.  Rodrigues-Vargas and A. Rueda (World
Scientific, Singapore, 1982), p.~428. See also P.W.  Milonni,
``Semiclassical and quantum-electrodynamical approaches in
nonrelativistic radiation theory'', {\sl Phys. Rep.} {\bf 25}, 1
(1976).

\item[{[5]}] A.V. Barranco, S.A. Brunini and H.M. Fran\c{c}a, ``Spin
and paramagnetism in classical stochastic electrodynamics'', {\sl Phys.
Rev.} {\bf A39}, 5492 (1989), and \ H.M.  Fran\c{c}a, T.W. Marshall and
E. Santos, ``Spontaneous emission in confined space according to
stochastic electrodynamics'', {\sl Phys.  Rev.} {\bf A45}, 6436 (1992).

\item[{[6]}] J. Dalibard, J. Dupont-Roc and C. Cohen-Tannoudji,
``Vacuum fluctuations and radiation reaction: identification of their
respective contributions'', {\sl J.  Phys.} {\bf 43}, 1617 (1982). See
also P.W. Milonni, ``Different ways of looking at the electromagnetic
vacuum'', {\sl Phys.  Scripta} {\bf T21}, 102 (1988).

\item[{[7]}] P.W. Milonni, ``The quantum
vacuum: an introduction to quantum electrodynamics'', Academic Press
Inc. (Boston, 1994), chapters 1, 2 and 8.

\item[{[8]}] A. Pais, ``Einstein and the quantum theory'', {\sl Rev.
Mod. Phys.} {\bf 51}, 863 (1979).

\item[{[9]}] A.V. Barranco and H.M. Fran\c{c}a, ``Einstein-Ehrenfest's
radiation theory and \linebreak Compton-Debye's kinematics'', {\sl
Found. Phys.  Lett.} {\bf 5}, 25 (1992); and ``Stochastic
electrodynamics and the Compton effect'', {\sl Physics Essays} {\bf 3},
53 (1990).

\item[{[10]}] See references [3] and [4] and also the papers by \ R.
Schiller and H. Tesser, ``Note on fluctuations'', {\sl Phys. Rev.} {\bf
A3}, 2035 (1971) \ and \ A.A. Sokolov and Tumanov, ``The uncertainty
relation and fluctuation theory'', {\sl Sov. Phys. (JETP)} {\bf 3}, 958
(1957).

\item[{[11]}] S. Bergia, P. Lugli and N. Zamboni, ``Zeropoint energy,
Planck's law and the prehistory of stochastic electrodynamics part 2:
Einstein and Stern's paper of 1913'', {\sl Ann. Found. Louis de
Broglie\/} {\bf 5}, 39 (1980).

\item[{[12]}] H.M. Fran\c{c}a and G.C. Santos, ``The extended charge in
stochastic electrodynamics'', {\sl Nuovo Cimento\/} {\bf 86B}, 51
(1985). See the Appendix.

\item[{[13]}] T.H. Boyer, ``Classical statistical thermodynamics and
electromagnetic zero-point radiation'', {\sl Phys. Rev.} {\bf 186}(5),
1304 (1969).

\item[{[14]}] S. Chandrasekhar, ``Stochastic problems in physics and
astronomy'', {\sl Rev. Mod. Phys.} {\bf 15}, 1 (1943). See Appendix IV,
p.~83.

\item[{[15]}] S. Bose, ``Planck's law and the light quantum
hypothesis'', {\sl Am. J. Phys.} {\bf 44}, 1056 (1976). This is an
English translation of the original 1924 paper.

\item[{[16]}] R. Blanco, H.M. Fran\c{c}a and E. Santos, ``Classical
interpretation of the Debye law for the specific heat of solids'', {\sl
Phys. Rev.} {\bf A43}, 693 (1991).

\item[{[17]}] D. Cole, ``Reviewing and extending some recent works on
stochastic electrodynamics'', in {\sl Essays on the Formal Aspects of
Electromagnetic Theory}, ed. A. Lakhtakia (1993), pp.~501--532. See
also D. Cole, ``Possible thermodynamic law violations and astrophysical
issues for secular acceleration of electromagnetic particles in
vacuum'', {\sl Phys. Rev.} {\bf E51}, 1663 (1995).

\item[{[18]}] M.J. Klein \ in \ {\sl ``Paul Ehrenfest''}, vol.~1
(North-Holland, 1985), chapter~10. See chapter 10, ``The essential
nature of the quantum hypothesis''.

\item[{[19]}] M.O. Scully and M. Sargent III, ``The concept of the
photon'', {\sl Phys. Today}, March 1972, p.~38.

\item[{[20]}] R. Kidd, J. Ardini and A. Anton, ``Evolution of the
modern photon'', {\sl Am. J. Phys.} {\bf 57}, 27 (1989); and ``Compton
effect as a double Doppler shift'', {\sl Am. J. Phys.} {\bf 53}, 641
(1985).

%\item[{[21]}] H.M. Fran\c{c}a and T.W. Marshall, {\sl Phys. Rev.} {\bf
%A38}, 6272 (1988).  See also K. Dechoum, H.M. Fran\c{c}a and A.M.
%Garcia, in ``Vacuum Fluctuations, Radiation Reaction and Sub-Heisenberg
%States'' (to be published).

\item[{[21]}] P. Grangier, G. Roger and A. Aspect, ``Experimental
evidence for a photon anti\-cor\-relation effect on a beam splitter: a
new light on single-photon interferences'', {\sl Europhys.  Lett.} {\bf
1}, 173 (1986).

\item[{[22]}] T.W. Marshall and E. Santos, ``Comment on {\em
Experimental evidence for a photon anticorrelation effect on a beam
splitter: a new light on single-photon interferences\/}'', {\sl
Europhys.  Lett.} {\bf 3}, 293 (1987); ``Stochastic optics: a
reaffirmation of wave nature of light'', {\sl Found. Phys.} {\bf 18},
185 (1988); ``The myth of the photon'', preprint, Univ. of Manchester
(October, 1995).

\item[{[23]}] R. Lange, J. Brendel, E. Mohler and W.  Martiensen,
``Beam splitting experiments with classical and with quantum
particles'', {\sl Europhys. Lett.} {\bf 5}, 619 (1988). See also J.
Brendel, S.  Sch\"utrumpf, R. Lange, W. Martienssen and M.O. Scully,
``A beam splitting experiment with correlated photons'', {\sl Europhys.
Lett.} {\bf 5}, 223 (1988); F. De Martini and S. Di Fonzo, ``Transition
from Maxwell-Boltzmann to Bose-Einstein partition statistics by
stochastic splitting of degenerate light'', {\sl Europhys. Lett.} {\bf
10}, 123 (1989).

\item[{[24]}] W. Nernst, ``An attempt to return, from the quantum
considerations, to the hypothesis of continuous changes in energy'',
{\sl Ver. Dsch. Phys. Ges.} {\bf 18}, 83 (1916).

\item[{[25]}] T.H. Boyer, ``Derivation of blackbody radiation spectrum
without quantum assump\-tions'', {\sl Phys. Rev.} {\bf 182}, 1374
(1969).

\item[{[26]}] T.H. Boyer, ``Derivation of the Planck radiation spectrum
as an interpolation formula in classical electrodynamics with classical
electromagnetic radiation'', {\sl Phys. Rev.} {\bf D27}, 2906 (1984).
See also the ``Reply to Comment on {\em Boyer's derivation of the Planck
spectrum\/}'', {\sl Phys. Rev.} {\bf D29}, 2477 (1984).

\item[{[27]}] T.H. Boyer, ``Derivation of the blackbody radiation
spectrum from the equivalence principle in classical physics with
classical electromagnetic zeropoint radiation, {\sl Phys. Rev.} {\bf
D29}, 1096 (1984).

\item[{[28]}] O. Theimer and P.R. Peterson, ``Statistics of classical
blackbody radiation with ground state'', {\sl Phys. Rev.} {\bf D10},
3962 (1974).

\item[{[29]}] M. Surdin, P. Braffort and A. Taroni, ``Black-body
radiation law deduced from stochastic electrodynamics'', {\sl Nature}
{\bf 23}, 405 (1966).

\item[{[30]}] J.L. Jim\'enez, L. de la Pe\~na and T.A. Brody,
``Zero-point term in cavity radiation'', {\sl Am.  J. Phys.} {\bf 48},
840 (1980).

\item[{[31]}] A.M. Cetto and L. de la Pe\~na, ``Continuous and discrete
aspects of blackbody radiation'', {\sl Found. Phys.} {\bf 19}, 419
(1989).

\item[{[32]}] A. Rueda, ``Behavior of classical particles immersed in
the classical zeropoint field'', {\sl Phys. Rev.} {\bf A23}, 2020
(1981); see section IV-B.

\item[{[33]}] R. Payen, ``Champs electromagn\'etiques al\'eatories:
formalisme g\'en\'eral et obtention de la loi de radiation de Planck'',
{\sl J. Physique\/} {\bf 45}, 805 (1984).

\item[{[34]}] R.H. Koch, D.J. Van Harlingen and J. Clarke,
``Observation of zero-point fluc\-tuations in a resistively shunted
Josephson tunnel junction'', {\sl Phys.  Rev. Lett.} {\bf 47}, 1216
(1981).

\item[{[35]}] S. Haroche and J.M. Raimond, ``Cavity quantum
electrodynamics'', {\sl Scient. Am.} {\bf 26} (April 1993).

\item[{[36]}] P.W. Milonni and M.L. Shih, ``Casimir forces'', {\sl
Contemporary Physics} {\bf 33}, 313 (1993). See also T.H. Boyer,
``Quantum zeropoint energy and long-range forces'', {\sl Ann. Phys.}
{\bf 56}, 474 (1970).

\item[{[37]}] Armin Hermann, ``The genesis of the quantum theory
(1899--1913)'', MIT Press (1971). See the Introduction and Chapter~1.

\end{list}

\end{document}